\documentclass[aps,prl,twocolumn,showpacs]{revtex4}
\usepackage{graphicx}
\usepackage{amssymb}
\usepackage{amsfonts}
\usepackage{amsmath}
\usepackage{dcolumn}

\def\Ea{{E_{\alpha}}}
\def\Ba{{B_{\alpha}}}
\def\Ca{{C_{\alpha}}}
\def\Va{{V_{\alpha}}}

\begin{document}

\title{First Principles Study of CaFe$_2$As$_2$ ``Collapse'' Under Pressure}
\author{Michael Widom}
\affiliation{Department of Physics, Carnegie-Mellon University, Pittsburgh, PA  15213}
\author{Khandker Quader}
\affiliation{Department of Physics, Kent State University, Kent, OH 44242}

\date{\today}

\begin{abstract}

We perform first principles calculations on CaFe$_2$As$_2$ under hydrostatic pressure. Our total energy calculations show that though the striped antiferromagnetic (AFM) orthorhombic (OR) phase is favored at P=0, a non-magnetic collapsed tetragonal (cT) phase with diminished c-parameter is favored for P $>$ 0.36 GPa, in agreement with experiments. Rather than a mechanical instability, this is an enthalpically driven transition from the higher volume OR phase to the lower volume cT phase. Calculations of electronic density of states reveal pseudogaps in both OR and cT phases, though As(p) hybridization with Fe(d) is more pronounced in the OR phase. We provide an estimate for the inter-planar magnetic coupling. Phonon entropy considerations provide an interpretation of the finite temperature phase boundaries of the cT phase.

\end{abstract}


\maketitle







Recent discoveries~\cite{Kamihara08} of iron-based pnictides provide a rich arena to explore the interplay between 
structural, magnetic and superconducting properties, and the consequent emergence of new physics. These materials provide insight into the competing roles of magnetism and pairing correlations, such as in the high temperature cuprate superconductors.
Amongst the pnictides, the 122 ternary compounds $A$Fe$_2$As$_2$ ($A$ = alkaline earth metal Ca, Ba, Sr), belonging to the  ThCr$_2$Si$_2$ structure family,
draw particular interest owing to the rich  behavior observed upon  chemical substitution or applying pressure~\cite{Alireza09, Mani09,Torikach09,Saha11}, such as different structural phases and superconductivity. Applied pressure has the advantage of introducing less disorder compared to chemical substitution. 

CaFe$_2$As$_2$, the smallest-volume member of this family, is of great current interest as it serves as a readily accessible system that exemplifies the key features of the $A$Fe$_2$As$_2$ compounds~\cite{Yu09,Canfield09}.
At ambient pressure, at T$_{c1}$ = 170 K, CaFe$_2$As$_2$ undergoes a 1st-order transition from a high temperature tetragonal (T) phase to a low temperature orthorhombic (OR) phase, that is striped along the $a$-axis and antiferromagnetically (AFM) ordered along the $c$-axis. This may be viewed as a magneto-structural transition from a high-T phase with fluctuating magnetic moments~\cite{Diallo10} to one with long-range AFM order. The striped magnetic order drives the orthorhombic symmetry breaking with the antiferromagnetic bonds in the $a$ direction being slightly longer than the ferromagnetic bonds in the $b$ direction.  The T-OR transition temperature $T_{c1}$ decreases with applied pressure.

At low-T, under hydrostatic pressure P $\sim$ 0.35 GPa~\cite{Canfield09}, the system undergoes a transition from the AFM-OR phase to a non-magnetic tetragonal phase, but with a compressed $c$-axis value; this has been termed the ``collapsed'' tetragonal phase (cT). At high-T, and P $>$ 0.35 GPa, another 1st-order transition occurs at
T$_{c2}$,  from the tetragonal T to the collapsed cT phase. T$_{c2}$ increases with pressure. Several features are sensitive to pressure conditions; in particular, lack of superconductivity up to P $\sim$ 0.65 GPa for the case of hydrostatic pressure~\cite{Yu09}, compared to observation of superconductivity under conditions creating uniaxial pressure~\cite{Kreyssig08,Torikach09}. Some experiments~\cite{Prokes10} indicate the presence of a low-T tetragonal phase sandwiched between the OR and the cT phase, suggesting that that superconductivity in a narrow region may be facilitated by the fluctuating moments present in the T phase. The transition from the OR to the cT phase occurs at lower pressures in the uniaxial case~\cite{Kreyssig08,Torikach09}.

Prior electronic density functional theory (DFT) work~\cite {Mazin10} has considered the pressure and doping dependence of BaFe$_2$As$_2$. More recently, DFT studies compared non-hydrostatic (i.e. anisotropic) and hydrostatic pressure dependences of BaFe$_2$As$_2$ and CaFe$_2$As$_2$~\cite{ Colonna11,Tomic12}, and proposed~\cite{Ji11} a Hund's rule coupling model of the phase transitions in these compounds.
Our DFT work provides a different understanding of CaFe$_2$As$_2$ under hydrostatic pressure, and goes beyond previous DFT work to incorporate a thermodynamic analysis of non-zero temperature and pressure. 

\begin{figure*}[ht!]
{\scalebox{0.30}{\includegraphics[clip,angle=0]{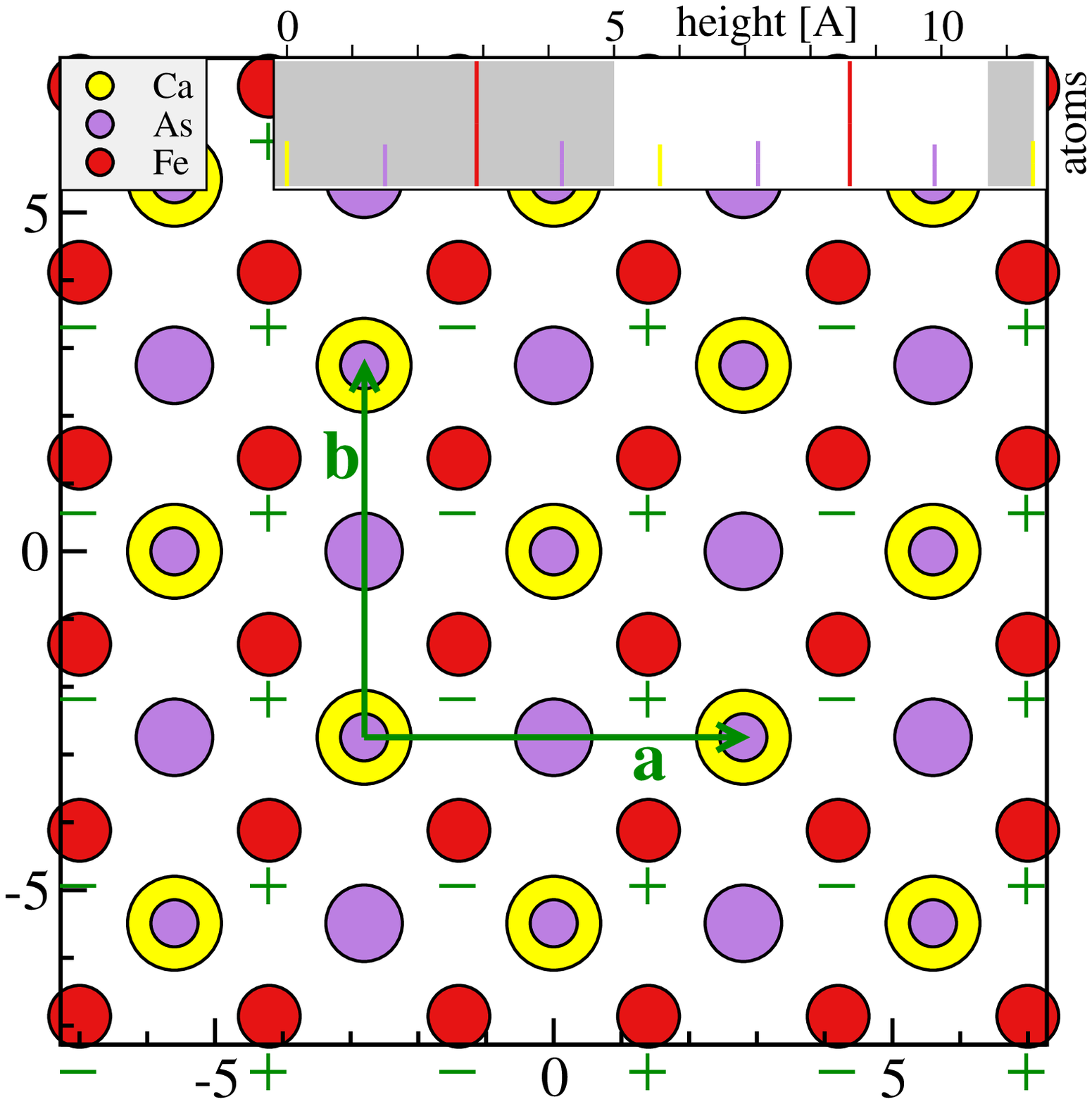}}}
{\scalebox{0.30}{\includegraphics[clip,angle=0]{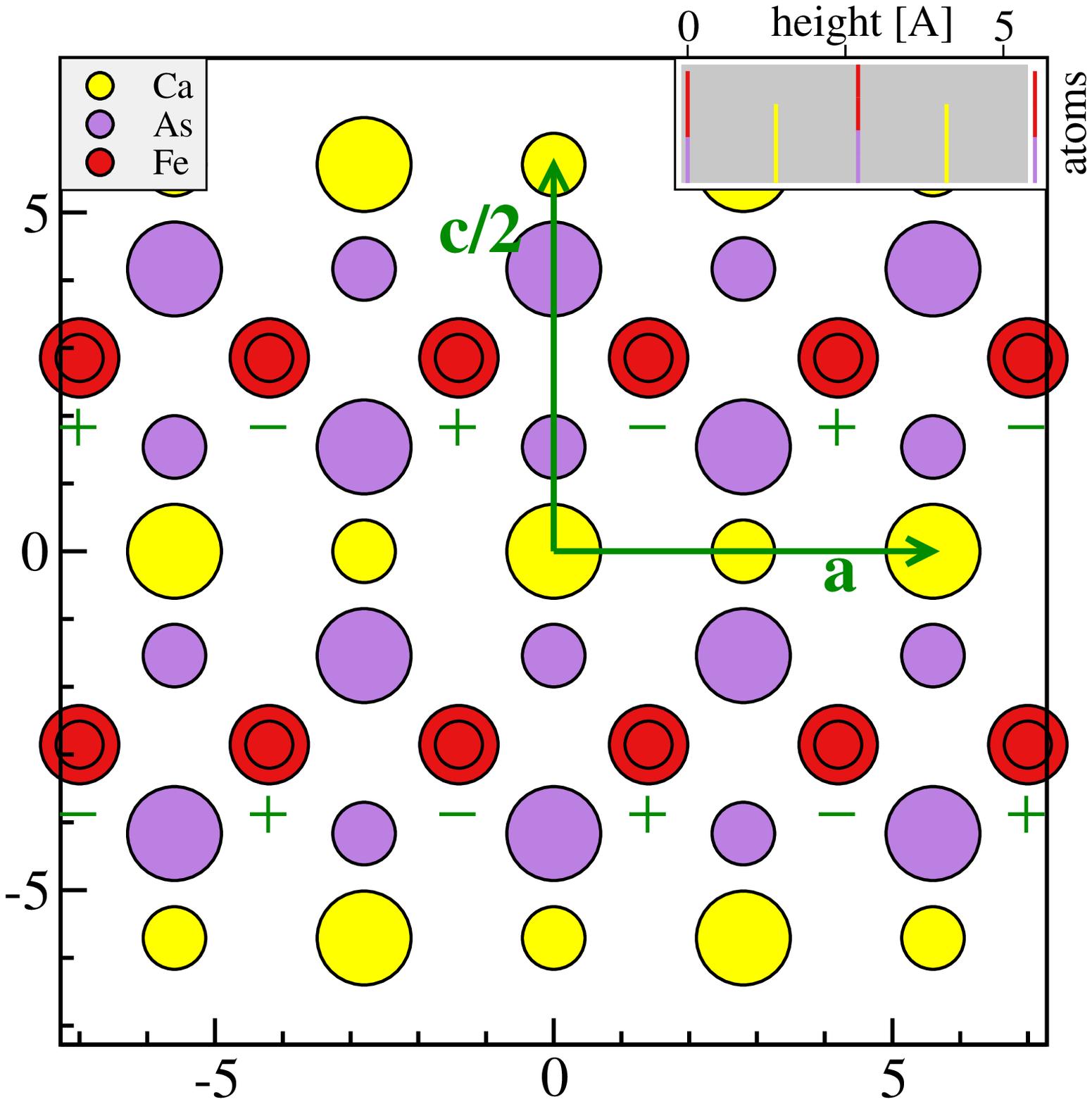}}}
\caption{(Color online) Structures used in calculations; chemical species are indicated by color, vertical heights are indicated by atomic size.
(a) Left panel shows the c-axis view,
with $ab$-plane stripe ordering denoted by +/-; (b) right panel shows the $b$-axis view with the $a$-axis stripe and the $c$-axis AFM ordering denoted by +/-.} 
\label{fig:struct}
\end{figure*}

The key results of our work are as follows.  Total energy considerations, as a function of pressure, lead to findings that agree well with experimental determination of electronic structure and magnetic ordering.  Though the higher volume AFM-OR phase  is favored at P = 0, the lower volume non-magnetic cT phase is favored for P $\ge$ 0.36 GPa.  Hence we assert the OR-cT transition is enthalpy-driven. Our density of states (DOS) calculations shows pseudogaps in both OR and the cT phases, with the pseudogap being narrower and deeper, and
As(p) hybridization with Fe(d) more pronounced, in the OR phase. Considering ferromagnetic (FM) and antiferromagnetic stripe orders, we provide an estimate of the inter-planar magnetic coupling. Finally, we show that thermodynamic calculations involving lattice phonons may provide useful insight into the observed phase boundaries at non-zero temperature and pressure.

The structures of interest are tetragonal, with Pearson symbol tI10 denoting body centered tetragonal with 10 atomic sites per unit cell, and orthorhombic, with Pearson symbol oF20 denoting face-centered orthorhombic with 20 atomic sites per unit cell.  The oF20 crystal structure is a based on a $\sqrt{2}\times\sqrt{2}$~R45 tetragonal supercell of the tI10 structure, followed by a weak orthorhombic distortion.  To improve consistency of our calculated property differences, we employ this tetragonal supercell for studies of the tetragonal structure, so we include 20 atoms in all our reported calculations.  Calculations of the orthorhombic structure utilize spin polarization, with initial moments in the striped antiferromagnetic arrangement.  Specifically, spins are ordered antiferromagnetically in the $a$ direction, ferromagnetically in the $b$ direction (i.e. ``striped'' in the $ab$-plane) and antiferromagnetically in the $c$ direction; 
see  Fig.~\ref{fig:struct}. The symmetry group of the atomic positions is $I4/mmm$ (no. 139) for tI10, and $Fmmm$ (no. 69) for oF20.

We utilize VASP~\cite{Kresse93,Kresse96} to carry out first principles total energy calculations, adopting projector augmented wave potentials~\cite{Blochl94,Kresse99}.  For a density functional we choose the PBE generalized gradient approximation, as calculations utilizing LDA fail to stabilize the striped antiferromagnetic orthorhombic phase, and the PW91 GGA predicts the orthorhombic phase to be energetically unstable at low temperature and pressure.  We relax all atomic positions and lattice parameters, and increase our k-point densities (to an 8x8x4 Monkhorst mesh) and plane-wave energy cutoff (to 340 eV) until energy differences have converged to 0.1 meV/atom.  Unusual care in selection of density functional and convergence is required because of the extremely small energy differences of order 1 meV/atom that must be resolved.

\begin{figure}[b]
{\scalebox{0.3}{\includegraphics[clip,angle=270]{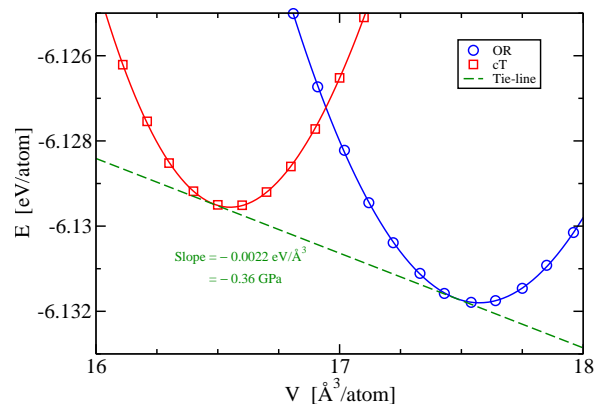}}}
\caption{(Color online) Calculated total energies (E) of the collapsed tetragonal (cT) and orthorhombic (OR) structures plotted against volume (V). The solid lines are cubic polynomial fits (see text; Table 1); the dotted line of double tangency gives the critical pressure beyond which cT is stable despite its higher energy.}
\label{fig:EvsV}
\end{figure}

Fig.~\ref{fig:EvsV} shows the results of our calculation of total energy vs. volume for the nonmagnetic cT and the AFM-OR phases, as described above. For each structure $\alpha$ we fit $\Ea(V)$ to a cubic polynomial of the form $\Ea=\Ea^0+(1/2\Va)\Ba(V-\Va)^2+(1/6\Va^2)\Ca(V-\Va)^3$, where $\Va$ and $\Ea^0$ are the volume and energy at $P=-dE/dV=0$, $\Ba$ is the bulk modulus and $\Ca$ is the nonlinear bulk modulus.  Fitted values of these quantities are listed in Table~\ref{tab:values}.  The line of double tangency $E_t=E^0-P_c^0 V$, where $P_c^0 = 0.022$ eV/\AA$^3$ (0.36 GPa) is the critical pressure.

\begin{table}[t]
\begin{tabular}{|r| lll| lll| l|}
\hline
Structure &  $b/a$  &  $c/a$ &  $\Va$  &  $E_\alpha^0$   &  $B_\alpha$  &  $C_\alpha$ & $m_{Fe}$ \\
\hline
cT    &   1     &  1.84  & 16.55 & -6.1296 & 85.6 & -654 &  0    \\
\hline
OR    &   0.98  &  2.02  & 17.57 & -6.1318 & 63.3 & -152 &  1.78 \\
\hline
\end{tabular}
\caption{\label{tab:values} Properties of structures at V=$\Va$.  Length and volume units are in \AA, energy in eV/atom, bulk moduli in GPa, and magnetic moment in Bohr magnetons.}
\end{table}

\begin{figure}[b]
{\scalebox{0.33}{\includegraphics[clip,angle=270]{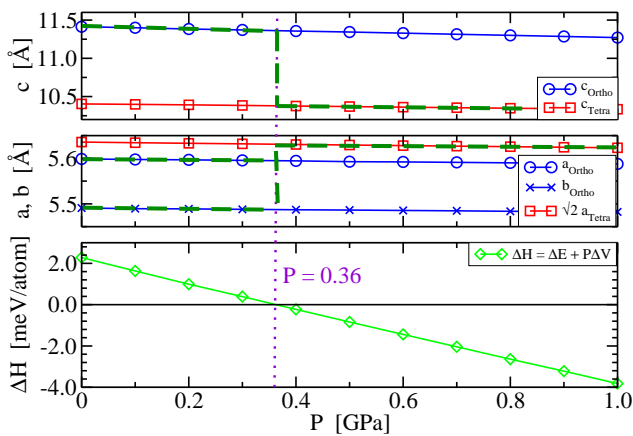}}}
\caption{(Color online) . The top two panels show the behavior of the calculated lattice parameters, $a,b,c$ of OR and cT structures with pressure. The bottom panel shows 
the change in enthalpy, $\Delta$H vs pressure, P.   Note the ``jumps'' in the lattice parameters at $P_c^0$ = 0.36 GPa (dotted lines), where $\Delta$H = 0 .} 
\label{fig:abcH}
\end{figure}

Fig.~\ref{fig:EvsV} implies that though the high volume OR phase is energetically favorable at low pressure, the enthalpies (H = E + PV) of the phases cross at $P_c^0$, beyond which the high energy cT phase, with diminished $c$-parameter, has lower enthalpy relative to that of the OR phase. This is more clearly seen in Fig.~\ref{fig:abcH} where we plot, for the OR and cT phases, the enthalpy difference ($\Delta$H=H$_{\rm cT}$-H$_{\rm OR}$), and the lattice parameters ($a,b,c$), versus pressure: While both the OR and the cT structures are stable across the pressure range studied, $\Delta$H vanishes at $P_c^0$ = 0.36 GPa, causing the c-axis lattice parameter to switch from the higher value of 11.45~\AA~ of the OR phase to a lower value of 10.45~\AA, the defining feature of the collapsed cT phase. Thus, the OR-cT transition at T=0 and finite-P is a transition in thermodynamic stability rather than a soft mode or irreversible mechanical instability as implied by the term ``collapse''.  In particular, the OR phase continues to exist at high pressure, and the cT phase is present at low pressure, they are simply metastable rather than thermodynamically stable.

Our calculated lattice parameters, shown in Fig.~\ref{fig:abcH}, across the pressure range (0 $\le$ P $\le$ 1 GPa) are within 2\% of experiments. At P$_c^0$ the discontinuities in lattice parameters are $\Delta$a=+0.04~\AA, $\Delta$b=+0.14~\AA, and $\Delta$c=$-$1.00~\AA, in good agreement with experiments~\cite{Kreyssig08}. On extending our calculations to higher pressures, P = 1.5 GPa, we do not find signatures for any other low-T transition.

\begin{figure}[t]
{\scalebox{0.35}{\includegraphics[clip,angle=270]{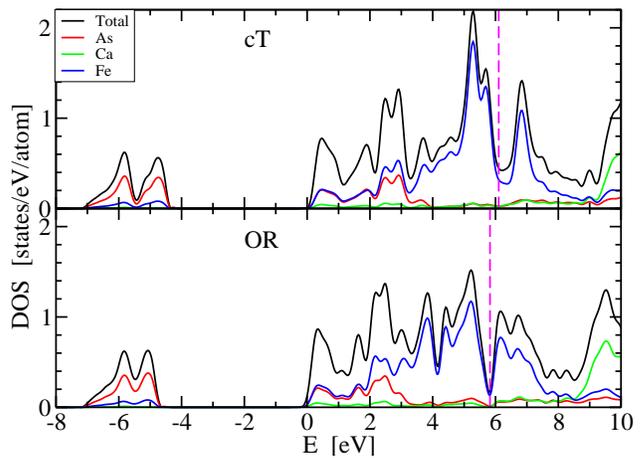}}}
\caption{(Color online) Calculated partial and total density of states (DOS) in cT (top panel) and OR (bottom panel) phases at P = 0.  $E_F$ is denoted by pink dashed lines.  Note the pseudogap features in OR and cT phases. As(p) hybridization with Fe(d) can be seen in OR phase.}
\label{fig:dos}
\end{figure}

In Fig.~\ref{fig:dos} we show our P = 0 calculated total electronic density of states (DOS), as well as the Fe, As, and Ca local DOS, for the OR and cT phases.  The As(p) hybridization with Fe(d) is found to be more pronounced in the AFM-OR phase than in the non-magnetic cT phase as can be seen in the As local DOS within 2 eV below $E_F$, and in the increased Fe $d$-band width over the same range, in the OR phase. Ca(s) orbital participation is insignificant in both cases. We have checked the s,p,d orbital-decomposed DOS and the band dispersion relations (not shown here) for consistency. This overall behavior persists for P$>$0. 
Both the OR and cT phases exhibit pseudogaps.  The Fermi level ($E_F$) lies in the center of the pseudogap in the OR phase, while the pseudogap is broader and shifted up from $E_F$ in the cT phase. We note that the  width and location of our DFT-calculated pseudogaps compare well
with LDA+DMFT calculations~\cite{Yin11} at P = 0.

As in other DFT calculations, our values of the Fe-moment at all pressures are large compared to the experimental results. At P = 0, the calculated $m_{Fe} \approx 1.78\; \mu_B$, compared to experimental value of 0.8 $\mu_B$. This decreases slightly with pressure. We estimate the inter-planar magnetic coupling based on our calculations of total energy and Fe magnetic moment, $m_{Fe}$.  The Heisenberg Hamiltonian for nearest-neighbor interactions is
\begin{equation}
\label{eq:magH}
H = - \sum_{<ij>}\;J_{ij}\;{\bf S}_i \cdot {\bf S}_j
\end{equation}
where $J_{ij}$ is the nearest-neighbor Heisenberg coupling between neighbors $i,j$, and have opposite signs for ferromagnetic and antiferromagnetic ordering, and ${\bf S_i}$ are the spin operators at site $i$. Under the usual assumption that the magnetic coupling is site-independent, the inter-planar magnetic coupling  $J_{\perp}$ is related to the difference in the total energy E of the stripe layers (Fig.~\ref{fig:struct}a) stacked ferromagnetically (FM) and antiferromagnetically (AFM, as shown in Fig.~\ref{fig:struct}b); i.e. $\Delta$E = E$_{\rm FM}$-E$_{\rm AFM}=2J_{\perp}\;m_{\rm Fe}^2$. Our DFT calculations give $\Delta$E=8.5 meV/atom.  Using the calculated  $m_{Fe} \approx 1.78\; \mu_B$ at P=0, we estimate $J_{\perp} \approx 2.7$ meV.  However, magnetization patterns with tetragonal symmetry (i.e. {\it un}striped along the $a$-axis, and also checkerboard in the $ab$-plane) lost their magnetization and relaxed to the cT state, illustrating the need to supplement eq.~(\ref{eq:magH}) with coupling of magnetism to lattice relaxation.

Although the first principles results are derived at T=0 K, we can estimate the temperature-dependent variation of critical pressure P$_c$(T) at low pressure to predict the finite-T phase boundary between the orthorhombic and collapsed tetragonal phases.  The Clausius-Clapeyron equation relates the slope of the phase boundary to the discontinuities in entropy  (S) and volume (V) as dP/dT=d$\Delta$S/d$\Delta$V. We calculated $\Delta$V $\equiv$ V$_{\rm cT}$ - V$_{\rm OR}\approx 1$~\AA$^3$/atom.  Temperature-dependent entropy can be calculated by integrating C/T, where C is the heat capacity.  We neglect the distinction between constant volume and constant pressure for low compressibility solids and assume that the dominant entropy contribution comes from the lattice phonons.  Note that at low-T the Debye approximation C = (12$\pi^4$/5)k$_B$(T/$\theta)^3$ becomes exact, where $\theta$ is the Debye temperature. Neglecting any temperature variation of $\theta$, the entropy also varies as (T/$\theta)^3$.  A value of $\theta_{\rm OR}$=292K has been reported experimentally~\cite{Ronning08}.  As we have not located an experimental value for $\theta_{\rm cT}$ we estimate $\theta_{\rm cT}\sim\left(B_{\rm cT}/B_{\rm OR}\right)^{1/2} \theta_{\rm OR}$=340K.  Note that $\Delta(\theta^{-3})\approx (-3\Delta\theta)/\theta^4$.

Approximating $\Delta$V and $\Delta\theta$ as constants allows us to integrate the Clausius-Clapeyron equation, yielding
\begin{equation}
\label{eq:Pc}
P_c(T)=P_c^0-\frac{3\pi^4}{5}
\left(\frac{k_B\Delta\theta}{\Delta V}\right)
\left(\frac{T}{\theta}\right)^4.
\end{equation}
Putting in our numerical values, $k_B\Delta\theta/\Delta V$=0.66 GPa, and $P_c^0$=0.36 GPa, we can invert Eq.~\ref{eq:Pc} to find the $T_c(P)$ phase boundary:
\begin{equation}
\label{eq:Tc}
T_c(P)=117K\times (P-0.36)^{1/4}
\end{equation}
in units of Kelvin.  Eq.~\ref{eq:Tc} implies that the boundary of the collapsed tetragonal phase rises vertically from its low temperature limit of 0.36 GPa, then bends sharply to the right towards higher pressures, in qualitative agreement with experiment~\cite{Yu09,Canfield09}.  The boundary curves to the right towards higher pressure because the higher bulk modulus of the cT phase reduces its entropy and hence raises its Gibbs free energy, G=E+PV-TS, relative to the OR phase.  At high temperature, it thus requires a higher pressure to favor the lower volume, but lower entropy, collapsed tetragonal phase.

Experimentally the collapsed tetragonal cT phase borders the orthorhombic OR phase at low temperatures, but it borders the non-collapsed tetragonal T phase at high temperatures.  In fact, the high temperature T phase resembles the OR phase in terms of its lattice parameters and even exhibits magnetic moment fluctuations matching the striped antiferromagnetic structure with correlation lengths of order 6-8~\AA~\cite{Diallo10}.  The T phase is essentially the OR phase with a loss of long-range order in magnetization leading to a loss of orthorhombicity (i.e. $a=b$), and hence the extrapolation of the cT-OR boundary to higher temperature and pressure should remain a good first approximation to the cT-T phase boundary~\cite{Kreyssig08}.

We thank A. Goldman for useful discussions and acknowledge the support of Aspen Center for Physics, where part of the work was carried out.

\bibliography{cafeas}

\end{document}